\documentclass[aps,preprint]{revtex4}
\usepackage[dvips]{graphics,graphicx}

\begin{document}
\title{Wannier-Stark states and Bloch oscillations in the honeycomb lattice}
\author{Andrey~R.~Kolovsky$^{1,2}$ and Evgeny~N.~Bulgakov$^{1}$}
\affiliation{$^1$Kirensky Institute of Physics, 660036 Krasnoyarsk, Russia}
\affiliation{$^2$Siberian Federal University, 660041 Krasnoyarsk, Russia}
%{\small \em andrey.r.kolovsky@gmail.com}
\date{\today}

\begin{abstract}
We study a quantum particle in a tilted honeycomb lattice in the tight-binding approximation. First we discuss the particle eigenstates, i.e., the stationary Wannier-Stark states. These states are proved to be extended states for the rational directions of the static field and localized states for the irrational directions. We find energy bands of the extended states and analyze the localized states. It is shown, in particular, that the localized `honeycomb' Wannier-Stark states are chaotic states with irregular dependence of the localization length on the static field magnitude. Second we discuss Bloch oscillations of the quantum particle. Irregular Bloch oscillations for irrational directions are observed.
\end{abstract}
%\pacs{05.60.Gg}{Quantum transport}
%\pacs{72.10.Bg}{General formulation of transport theory}
%\pacs{05.45.-a}{Nonlinear dynamics and chaos}
\maketitle

%%%%%%%%%%%%%%%%%%%%%%%%%%%%%%%%%
\section{Introduction}

Wannier-Stark states (WS-states) are eigenstates of quantum particle in a tilted lattice. Strictly speaking WS-states are resonances and have complex energies. However, for  weak static  fields they can be approximated by the stationary states with real energies (the single-band  or tight-binding approximations). In the past two decades WS-states and related problems of Bloch oscillations (BO) and interband Landau-Zenner tunneling (LZ-tunneling) were readdressed in a number of fascinating laboratory experiments with cold atoms in (quasi) 1D optical lattices  and the light in 1D arrays of optical waveguides, see Refs.~\cite{Daha96,Raiz97,Mors01,Kling10, Pert99,Mora99,Trom06b,Drei09} to cite a few of dozens relevant papers.  These experiments stimulated theoretical studies that resulted in essential progress in the theory of WS-states in one-dimentional systems, see Ref.~\cite{53} for a review.

An interesting extension of the theory of 1D WS-states refers to 2D tilted lattices \cite{51,58,Witt04}. It was argued in Ref.~\cite{51} and confirmed later on in the experiment \cite{Trom06b} that WS-states in a 2D lattice are sensitive to the direction of the static field relative to primary axes of the lattice. Unfortunately, for square lattices considered in the above cited papers this effect is well pronounced only in the strong-field regime, where metastable nature of WS-states plays major role. In the present work we analyze WS-states and BO for a quantum particle in a honeycomb lattice. We will show that for this lattice the non-analitic angular dependence of WS-states is seen already in the weak-field regime, where the metastable WS-states can be approximated by the stationary states. This feature of the stationary `honeycomb' WS-states  has direct consequences for Bloch dynamics of the system that becomes qualitatively different for rational and irrational field directions defined later on in the text.  It should be mentioned that BO in the honeycomb lattice were addressed earlier with respect to conductivity of graphene nano-ribbons \cite{Ferr11,Krue12}. However, in these works the electric field was aligned with the ribbon axis and, thus, the alignment effects were not discussed.

This paper consists of two parts devoted to analytical and numerical analysis of WS-states,  Sec.~\ref{sec2}, and BO of a quantum particle in the honeycomb lattices, Sec.~\ref{sec3}. 
%Since these two problems are the spectral and dynamical aspects of one Wannier-Stark problem, approaches of Sec.~\ref{sec2} and Sec.~\ref{sec3} accomplish each other.  
The main results of the work are summarized in Sec.~\ref{sec4}.

%%%%%%%%%%%%%%%%%%%%%%%%%%%%%%%%%
\section{Honeycomb Wannier-Stark state}
\label{sec2}

In the standard presentation with two sublattices  the tight-binding Hamiltonian of a quantum particle in the honeycomb lattice reads
\begin{equation}
\label{0}
H_0= -J\sum_{\bf R} \left(\sum_{j=1}^3 b^\dagger_{{\bf R}+{\bf r}_j} a_{{\bf R}} +h.c.\right) \;,
\end{equation}
where $J$ is the hopping matrix element, ${\bf R}$ denote coordinates of A-sites, and ${\bf r}_j$ are three vectors that point from a A-site to the nearest B-sites. If a static field is present, this Hamiltonian should be complimented with the Stark term,
\begin{equation}
\label{1}
H= H_0
+\sum_{{\bf R}} ({\bf F}, {\bf R})   a^\dagger_{{\bf R}}  a_{{\bf R}}
+\sum_{{\bf R}'} ({\bf F} ,{\bf R}') b^\dagger_{{\bf R}'} b_{{\bf R}'} 
\end{equation}
where ${\bf F}$ is the field vector and ${\bf R}$ and ${\bf R}'$ are coordinates of the A- and B-sites, respectively. We are interested in the eigenstates of the Hamiltonian (\ref{1}), i.e., in the stationary honeycomb WS-states.

We begin with recalling general result concerning the structure of WS-states in a two-dimensional lattice of arbitrary geometry: For the field  ${\bf F}$ parallel to a vector pointing from one lattice site to any other site WS-states are extended states in the direction orthogonal to ${\bf F}$. These field directions can be labeled by two co-prime numbers $r$ and $q$ and for this reason are turmed rational directions. For example, for the square lattice the rational directions are given by $F_y/F_x=r/q$ or
\begin{equation}
\label{2}
\tan\theta=r/q \;,
\end{equation}
while for the honeycomb lattice these are
\begin{equation}
\label{3}
\tan\theta=\sqrt{3}\frac{q-r}{q+r} \;.
\end{equation}
The spectrum of WS-states for a rational direction $(r,q)$  consists of infinite number of equally spaced energy bands. A particular feature of the square lattice and other simple lattices (like, for example, a triangular lattice with the hexagonal symmetry \cite{remark1}) is that these energy bands are flat for almost all rational directions. Because of this feature there is no qualitative difference between rational and irrational directions for the square lattice in the tight-binding approximation. In what follows we show that the Wannier-Stark energy bands of the honeycomb lattice have finite widths already in the tight-binding approximation. This makes a crucial difference between the honeycomb lattice and the square or triangular lattices, as well as between rational and irrational directions for the honeycomb lattice.

%%%%%%%%%%%%%%%%%%%%%%%%%%%%%%%%%%%%%%%%%%%
\subsection{Rational field directions}

For the rational directions WS-states in the honeycomb lattice are labeled by the ladder number $n$, the transverse quasimomentum $\kappa$, and the sublattice index $i$. Correspondently, the energy spectrum is given by
\begin{equation}
\label{b1}
E_n^{(i)}(\kappa)=E_0^{(i)}+d \tilde{F} n+\epsilon^{(i)}(\kappa) \;,
\end{equation}
where 
\begin{equation}
\label{b2}
d=\frac{1}{ \sqrt{r^2+q^2}} \;,\quad
\tilde{F}=\frac{3F}{2d}\frac{1}{\sqrt{r^2-rq+q^2}}  \;,
\end{equation}
and $\epsilon^{(i)}(\kappa)$ is a periodic function of $\kappa$. 
%We calculated the spectrum (\ref{b1}) by using the method described in detail in Appendix A. 
We calculated the spectrum (\ref{b1}) by adopting the method of Ref.~\cite{Naka93,Naka95,preprint1}. In brief, we map the honeycomb lattice into a square lattice with two sublattices. For this square lattice the static field is characterized by the vector $\tilde{{\bf F}}$ and the rational directions are given by $\tilde{F}_y/\tilde{F}_x=r/q$. Next we introduce another square lattice of the period $d$, which includes the previous lattice as a sublattice, and rotate it to align its $x$ axis with the vector $\tilde{{\bf F}}$. Finally we use substitution where the wave function is a plane wave along the $y$ axis. After this sequence of transformations we end up with the system of two coupled 1D equations,
\begin{eqnarray}
\label{b3}
-J(e^{-iq \kappa d}\psi^B_{j-r} + e^{ir \kappa d}\psi^B_{j-q} + e^{i(r-q) \kappa d}\psi^B_{j-r-q})
+ (d\tilde{F} j + E_0^A) \psi^A_j= E\psi^A_j  \;, \\
\nonumber
-J(e^{iq \kappa d}\psi^A_{j+r} + e^{-ir \kappa d}\psi^A_{j+q} + e^{i(q-r) \kappa d}\psi^A_{j+r+q})
+ (d\tilde{F} j +E_0^B)  \psi^B_j= E\psi^B_j  \;,
\end{eqnarray}
where $E_0^B=(2/3)(r+q)d\tilde{F}$ if one sets $E_0^A=0$.
%#############################################
\begin{figure}
\center
\includegraphics[height=9cm,clip]{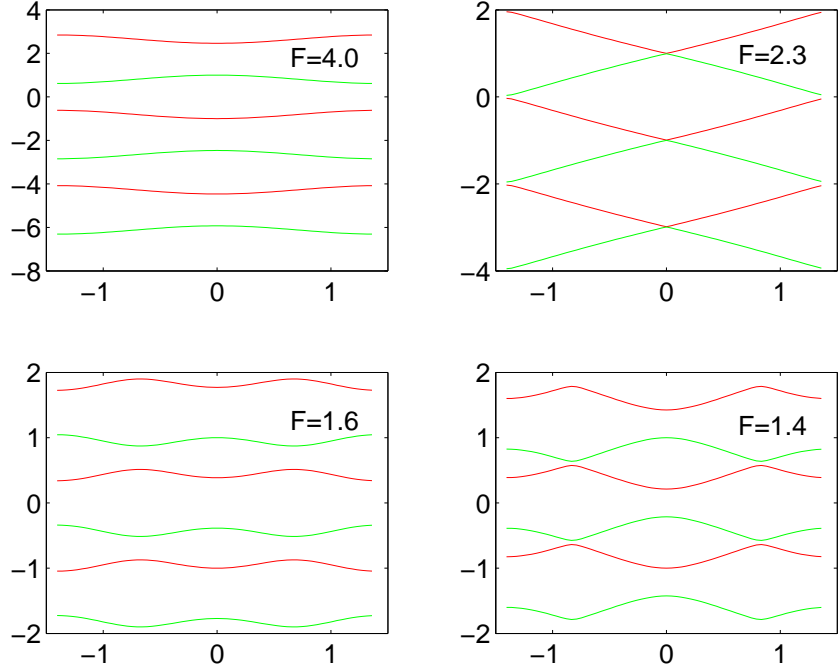}
\caption{Fragments of the spectrum (\ref{b1}) for $(r,q)=(1,2)$ or $\theta=\pi/6$ and different values of the field magnitude $F$.}
%$F=2.582,1.6,1.05,0.2$.
\label{fig5}
\end{figure}
%#############################################
\begin{figure}
\center
\includegraphics[height=9cm,clip]{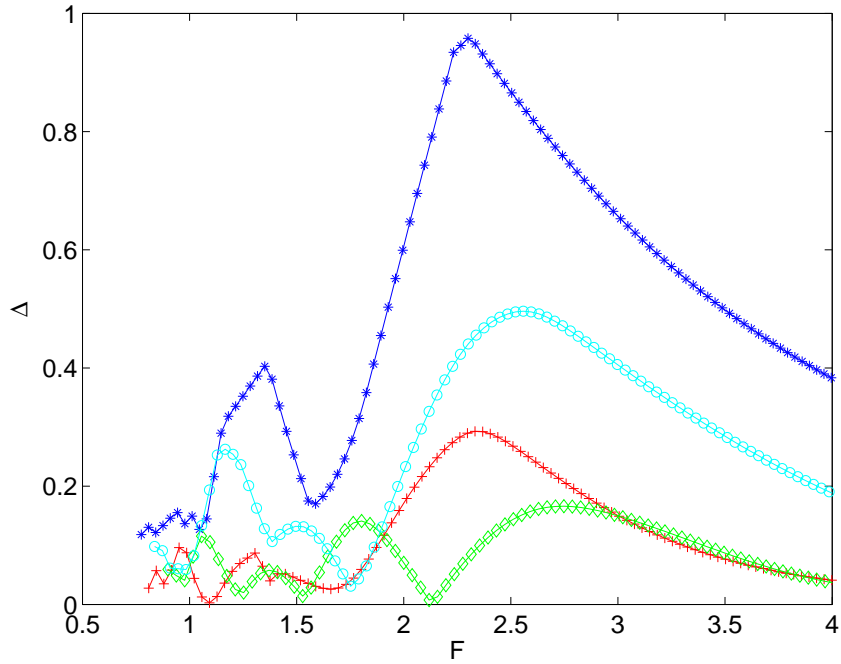}
\caption{The band width (\ref{b4}) as the function of the field magnitude for $(r,q)=(1,2)$, asterisks, $(1,3)$, circles, $(1,5)$, diamonds, and $(2,5)$, crosses.}
\label{fig6}
\end{figure}

We solved Eq.~(\ref{b3}) numerically for different $F$ and $(r,q)$. Without any lost of generality  one can restrict $\theta$ to the interval $0\le\theta<\pi/3$ which means $q>r\ge0$. Examples of the spectrum (\ref{b1}) are given in Fig.~\ref{fig5} where the chosen energy intervals include three band with $i=1$ and three bands with $i=2$.  In addition to Fig.~\ref{fig5}, Fig.~\ref{fig6} shows the width of the energy bands,
\begin{equation}
\label{b4}
\Delta=\max_\kappa \epsilon(\kappa) - \min_\kappa \epsilon(\kappa)  \;,
\end{equation}
as the function of $F$ for the field direction $\theta=\pi/6$ (asterisk) and some other directions. [Since $\Delta$ is independent of $i$, we drop the sublattice index in Eq.~(\ref{b4}).]  From the depicted numerical data one draws the following conclusions: The bands are well separated  only in the limit of large $F$; With decrease of $F$ the band width $\Delta$ monotonically grows while the distance $d\tilde{F}$ between bands monotonically decreases and, for some $F_{cr}\sim J$, the bands almost touch each other. At this critical field magnitude $\Delta$ takes its maximal value; After reaching the maximum $\Delta$ shows monotonic decrease, where bands become flatter; This decrease is followed by erratic oscillations of the band width for small $F$. We note that in this region of small $F$ the quantity (\ref{b4}) is not sufficient to  characterize the spectrum because of rather complicated band pattern with many avoided crossings. 

We also studied the asymptotic behavior of $\Delta$ for $F\rightarrow\infty$. Our numerical analysis reveals the dependence
\begin{equation}
\label{b5}
\Delta \sim \frac{1}{F^\nu}  \;,
\end{equation}
where $\nu$ increases with the increase of $(r,q)$. For example, $\nu=1$ for $(r,q)=(1,1)$, $\nu=2$ for $(r,q)=(1,2)$, etc. This result is similar to that for the band widths of the Landau-Stark states (eigenstates of a quantum particle in the Hall configuration) in the square lattice \cite{preprint1}. We believe that the power $\nu=\nu(r,q)$ can be calculated  analytically by adopting the perturbative approach of Ref.~\cite{preprint1}.
%#############################################
\begin{figure}
\center
\includegraphics[height=9cm,clip]{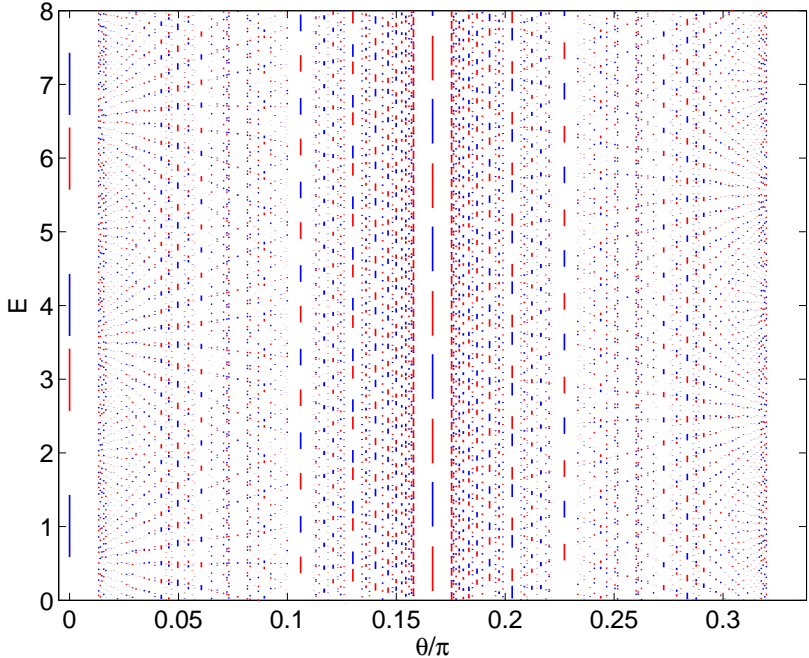}
\caption{Widths of the energy bands for different rational directions for $F=2$.}
\label{fig7}
\end{figure}

Dependence of the band width $\Delta$ on the field direction $\theta$ is depicted in Fig.~\ref{fig7}, which presents the energy spectrum of the system (\ref{1}) in the form of Hofstadter's butterfly. Namely, the figure shows energy bands for angles (\ref{3}) with $1<r\le q\le 21$, where the blue (dark grey) and green (light grey) colors correspond to $i=1$ and $i=2$, respectively. 
%A number of  symmetries, which are discussed in detail in Appendix B, are noticed. 
It is seen in Fig.~\ref{fig7} that the band widths progressively decrease with increase of $(r,q)$. Together with the estimate (\ref{b5}) this means that for irrational directions the spectrum is pure point and, hence, WS-states are localized states.

%%%%%%%%%%%%%%%%%%%%%%%%%%%%%%%%%%%%%%%%%%%
\subsection{Irrational field directions}

\begin{figure}[b]
\center
\includegraphics[width=12cm,clip]{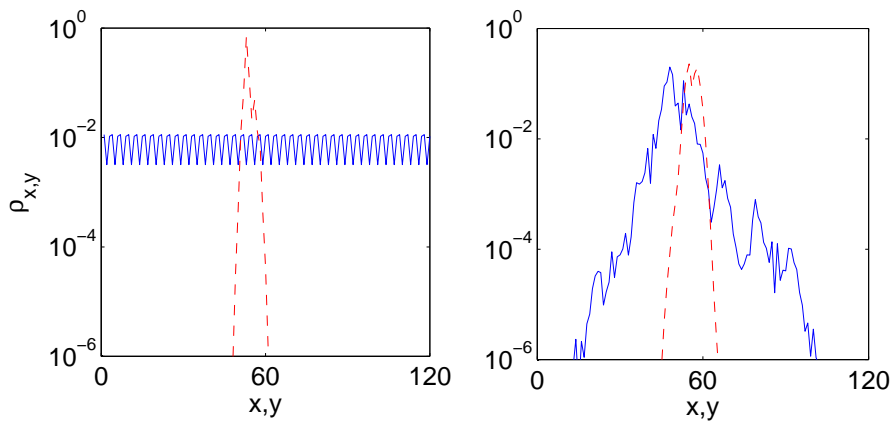}
\caption{Examples of extended (left panel, $\theta=\pi/2$) and localized (right panel,  $\theta=3-\pi/2$) WS-states with nearly the same energies. The dashed and solid lines show integrated probabilities along and across the field, respectively. The field magnitude $F=1$.}
\label{fig3}
\end{figure}

First we check that WS-states for irrational directions are localized states. In Fig.~\ref{fig3} we compare two eigenstates of the Hamiltonian (\ref{1}) with nearly the same energy for a `rational' $\theta=\pi/2\approx 1.57$ (equivalent to $\theta=\pi/6$)  and `irrational' $\theta=3-\pi/2\approx 1.43$.  The figure shows the integrated probabilities $\rho_y=\int |\Psi({\bf R})|^2 {\rm d}x$, dashed line, and $\rho_x=\int |\Psi({\bf R})|^2 {\rm d}y$, solid line. (From now on we do not distinguish A- and B-sites.) For rational $\theta$ the WS-state is seen to be an extended state in the direction orthogonal to ${\bf F}$, while for irrational $\theta$ it is  localized in both directions.
%#############################################

An important characteristic of the localized WS-states is their participation ratio,
\begin{equation}
\label{b6}
P=\left(\sum_{\bf R} |\Psi({\bf R})|^4\right)^{-1} \;,
\end{equation}
which indicates how many sites are occupied by a given state.  We note that the participation ratio (\ref{b6}) is the same for any WS-state because different WS-states are related to each other by translations. In the other words, one can obtain the complete set of WS-states from one state (or few states, if the lattice consists of sublattices) by translating it across the lattice.

\begin{figure}
\center
\includegraphics[height=9cm,clip]{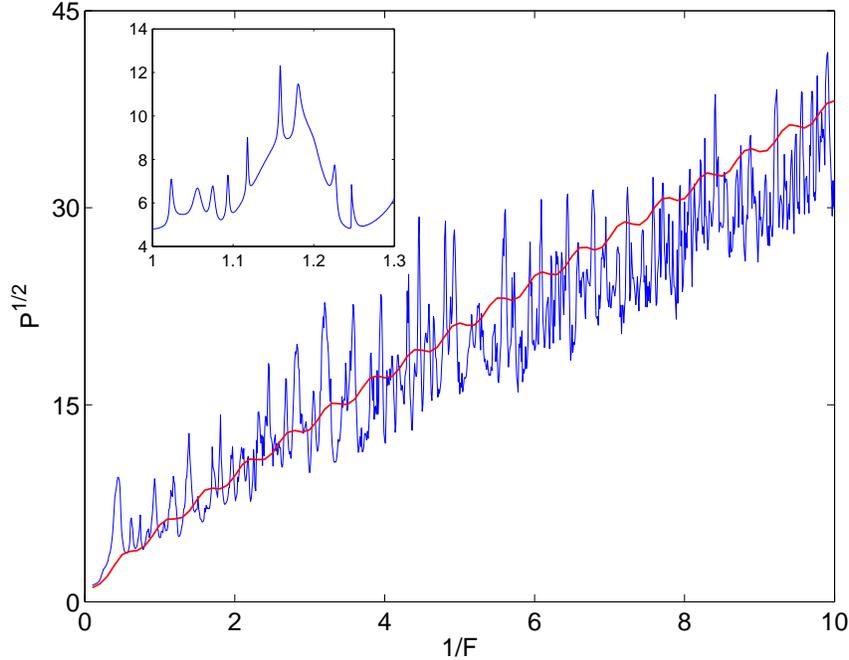}
\caption{Localization length of WS-state for irrational direction $\theta=3-\pi/2$. The thick and thin lines show the square root of the participation ratio (\ref{b6}) of WS-states for the square and honeycomb lattices, respectively. The inset zooms into the region $1/F\approx 1$.}
\label{fig4}
\end{figure}

A remarkable feature of the honeycomb WS-states is that their participation ratio wildly oscillates if $F$ is varied. The physics behind these oscillations is the following. Similar to the case considered in the previous subsection the (now discrete) spectrum of WS-state consists of two subsets that can be labeled by the subband index $i$ or, what is the same, by the letters  A and B. Correspondently, we have two families of WS-states \cite{remark4}.  When $F$ is varied the energy levels of A- and B-states non-monotonically move on the energy axis.  If two levels of different symmetry come close to each other  they develop an avoided crossing where the A- and B-states hybridize. As a consequence of the hybridization the function $P=P(F)$ shows a local maximum, see inset in Fig.~\ref{fig4}. It should be mentioned that to resolve all local peaks of $P(F)$ (i.e, all avoided crossings in the spectrum) the step over $F$ or, more precisely, over $z=1/F$ should be infinitesimally small. Figure \ref{fig4} shows the function $P=P(F)$ for a moderate step where only large peaks are resolved. Erratic oscillations with increasing density of peaks are clearly seen. This figure also reveals the expected average growth of the participation ratio when $F$ is decreased.
%#############################################

It is interesting to compare the localized honeycomb WS-states against the analytic results for the simple square lattice, 
\begin{eqnarray}
\label{b7}
\Psi({\bf R})={\cal J}_{l-n}(J/F_x) {\cal J}_{m-k}(J/F_y) \;,\quad {\bf R}=(l,m) 
\end{eqnarray}
(here ${\cal J}_n(z)$ is the Bessel functions of the first kind). The square lattice has only one family of WS-states and, correspondently, the energy levels show no avoided crossings. Participation ration of the states (\ref{b7}) is depicted in Fig.~\ref{fig4} by the thick line. Comparing two curves we conclude that  WS-state in the honeycomb lattice are irregular or chaotic states, which are known to be sensitive to variations of the system parameters. The statistical analysis of these states in spirit of the Random Matrix Theory will be presented elsewhere.

%%%%%%%%%%%%%%%%%%%%%%%%%%%%%%%%%%
\section{Bloch dynamics}
\label{sec3}

To study Bloch dynamics of the system (\ref{1}) it is convenient to use the interaction representation with respect to the Stark term. This results in the time-dependent Hamiltonian
\begin{equation}
\label{4}
H(t)= -J\sum_{\bf R} \left(\sum_{j=1}^3 b^\dagger_{{\bf R}+{\bf r}_j} a_{{\bf R}}
e^{i\omega_j t} +h.c.\right)
\end{equation}
where $\omega_j=({\bf F},{\bf r}_j)$ are the Bloch frequencies. Notice that the Hamiltonian (\ref{4}) commutes  with the translation operator. Thus, when considering translation-invariant solutions of the Schr\"odinger equation, we can impose periodic boundary conditions. 

%%%%%%%%%%%%%%%%%%%%%%%%%%%%%%%%%
\subsection{Delocalized initial state}

\begin{figure}
\center
\includegraphics[height=9cm,clip]{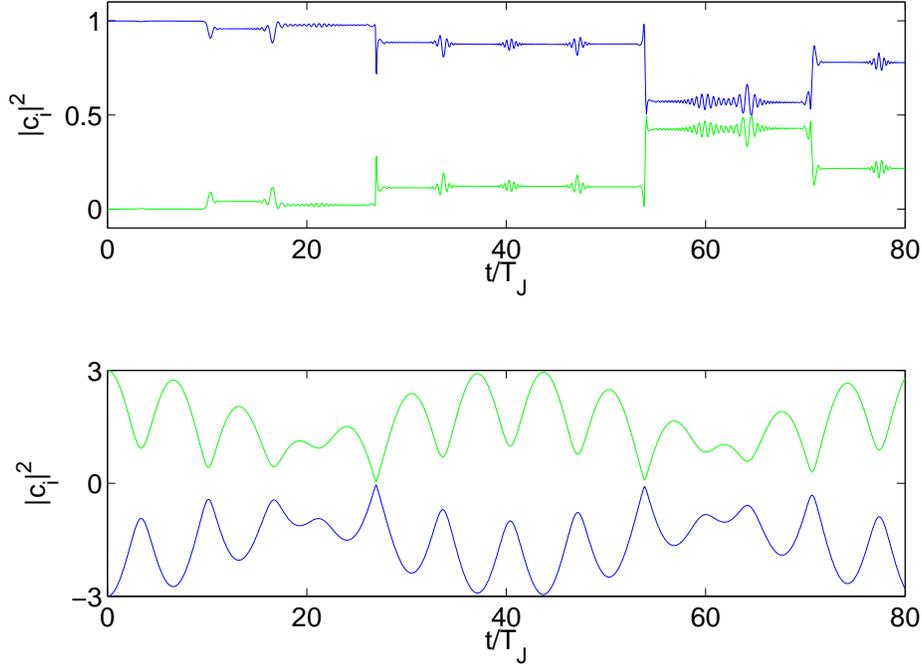}
\caption{Populations of two Bloch subbands as the functions of time, upper panel, and the energies (\ref{5}) for ${\bf k}'={\bf k}+{\bf F}t$, lower panel. Parameters are $\theta=\pi-3$ and $F=1$.}
\label{fig9}
\end{figure}

Consider initial state of the system $\Psi({\bf R},t=0)$ given by an eigenstate of the Hamiltonian (\ref{0}), i.e., by the Bloch wave $\Psi_{\bf k}({\bf R})$ with the quasimomentum ${\bf k}$. Then the time evolution of this state is naively expected to obey the equation $\Psi({\bf R},t)\sim \Psi_{{\bf k}+{\bf F}t}({\bf R})$, that is known as the acceleration theorem. This simple Bloch dynamics, however, is complicated by the fact that the spectrum of $H_0$ consist of two subbands,
\begin{equation}
\label{5}
E({\bf k})=\pm\sqrt{1+4\cos^2\left(\frac{\sqrt{3}}{2} k_y \right)
 + 4\cos\left(\frac{\sqrt{3}}{2} k_y \right)\cos\left(\frac{3}{2} k_x \right) } \;,
\end{equation}
with the energy gap vanishing at the Dirac points. Thus one generally can not avoid interband LZ-tunneling and the actual time evolution is given by the equation
\begin{equation}
\label{6}
\Psi({\bf R},t)=\sum_{i=1}^2 c_i(t) \Psi^{(i)}_{{\bf k}+{\bf F}t}({\bf R}) \;,
\end{equation}
where $i$ is the Bloch subband index. The upper panel in Fig.~\ref{fig9} shows a typical dynamics of the coefficients $c_i(t)$ and the lower panel depicts the energies (\ref{5}) at ${\bf k}'={\bf k}+{\bf F}t$. It is seen that LZ-tunneling predominantly  takes place when ${\bf k}'$ appears in the vicinity of the Dirac points. 
%#############################################
%#############################################
\begin{figure}
\center
\includegraphics[height=9cm,clip]{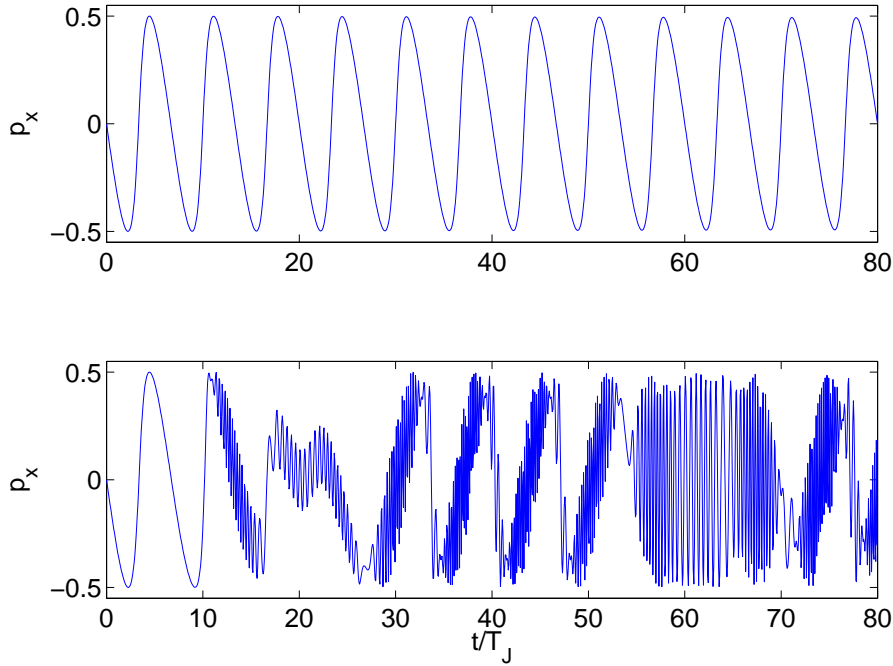}
\caption{Projection of the mean momentum on the $x$ axis as the function of time for $F=1$ and $\theta=0$, upper panel, and $\theta=\pi-3$, lower panel.}
\label{fig8}
\end{figure}

We would like to mention that the discussed Landau-Zener transitions may completely smear the periodic or quasiperiodic dynamics of quantum observables that is usually associated with BO. Figure \ref{fig8} compares dynamics of $p_x$, the projection of the mean momentum on the $x$ axis, for $F=1$ and $\theta=0$ (upper panel) and $\theta=\pi-3$ (lower panel). In the former case the particle trajectory in the (quasi)momentum space goes between Dirac points and LZ-tunneling can be neglected for chosen $F$. Thus the upper subband remains unpopulated and we observe nice periodic oscillations of $p_x$. In the latter case the upper subband gets populated independent of how  small $F$ is. As a consequence, $p_x$ shows irregular oscillations where one hardly recognize the former periodic BO.

%%%%%%%%%%%%%%%%%%%%%%%%%%%%%%%%%
\subsection{Localized initial state}

Wave-packet dynamics in a 2D lattice with two Bloch subbands was considered  earlier in Ref.~\cite{58,Witt04}. The wave packet was found to have a tendency to spread in the direction orthogonal to ${\bf F}$, while in the direction parallel to  ${\bf F}$ it shows oscillatory dynamics.  In this work we consider the limiting case of a localized initial packet where only one site is populated at $t=0$.  Also we will discuss dynamics in terms of WS-states instead of discussing it in terms of Bloch states as it was done in the above cited papers. To describe the wave-packet dynamics we introduce the time-dependent analogue of Eq.~(\ref{b6}),
\begin{equation}
\label{7}
P(t)=\left(\sum_{\bf R} |\Psi({\bf R},t)|^4\right)^{-1} \;.
\end{equation}
According to the results of Sec.~\ref{sec2} we expect qualitatively different dynamics of the participation ratio (\ref{7}) for rational and irrational field directions. 

\begin{figure}
\center
\includegraphics[height=9cm,clip]{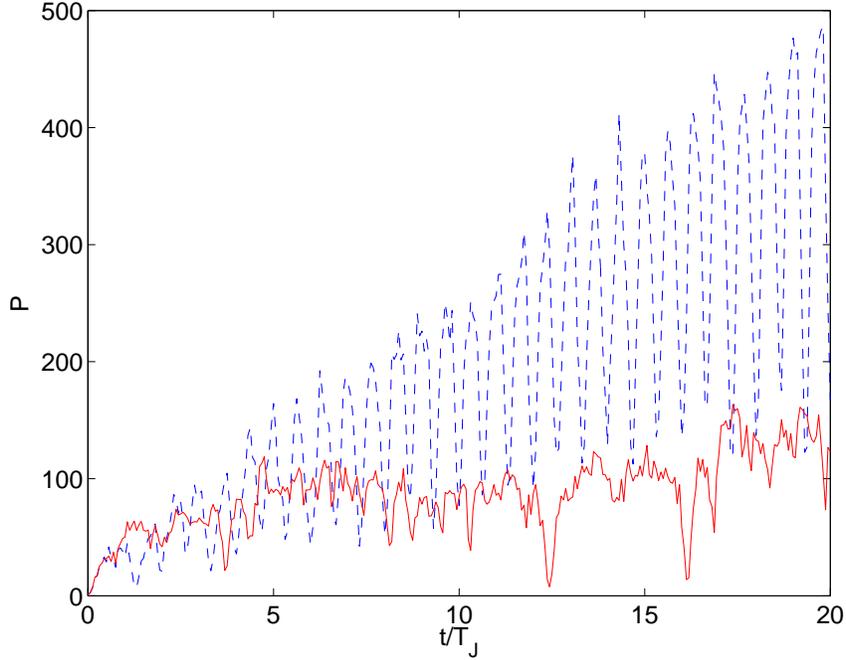}
\caption{Participation ratio (\ref{7}) for $\theta=0$ (dashed line) and $\theta=\pi-3$ (solid line). The initial condition corresponds to population of a single site in the center of the lattice. The value of the static force $F=1$.}
\label{fig1}
\end{figure}

The dashed line in Fig.~\ref{fig1} shows $P(t)$ for $F=1$ and rational direction $\theta=0$. The participation ratio exhibits oscillatory dynamics superimposed with a linear increase in the mean value. This linear increase is due to ballistic spreading of the packet in the direction orthogonal to ${\bf F}$. The rate of ballistic spreading is obviously defined by the width $\Delta$ of the Wannier-Stark bands, while the characteristic frequency of oscillations is given by the distance between neighboring Wannier-Stark bands. 
%#############################################

The case of irrational direction $\theta=\pi-3$ is depicted by the solid line in Fig.~\ref{fig1}. Now $P(t)$ saturates at some level defined by the characteristic localization length of the honeycomb WS-states. The wave-packet simulations also confirm $1/F^2$ scaling law for saturation level of $P(t)$ that follows from the $1/F^2$ scaling law for the participation ratio of the localized honeycomb WS-states.

We would like to mention that for some choices of the system parameters we observed rather exotic wave-packet dynamics. One example is $\theta=\pi/6$ and $F=2.3$. As it is seen from Fig.~\ref{fig5}(b),   for these parameters the energy bands are almost straight lines, that implies nondispersive dynamics. Indeed, in our simulations we observed the formation of two soliton-like packets that propagate with constant velocities in opposite directions.

%%%%%%%%%%%%%%%%%%%%%%%%%%%%%%%%%%
\section{Conclusions}
\label{sec4}

We  found WS-states of a quantum particle in a tilted honeycomb lattice and compared them with WS-states in the tilted square lattice that are known analytically.  The comparison is done for both rational and irrational directions of the  field vector ${\bf F}$. %Both similarities and differences are found.

For rational directions of the field defined in Eqs.~(\ref{2}-\ref{3}) the energies of WS-states form energy bands. For the square lattice these bands have zero width, excluding the case where  ${\bf F}$ is aligned with one of two primary axes. This prohibits any transport in the system if the vector ${\bf F}$ is misaligned with a primary axis.  Unlike this situation, for the honeycomb lattice the bands have finite width for any rational direction. If the condition (\ref{3}) is fulfilled, an initially localized wave packet spreads in the direction orthogonal to ${\bf F}$.  It is interesting to notice that this type of wave-packet dynamics exist in the square lattice only in the presence of a gauge field normal to the lattice plane \cite{preprint1}.

For irrational directions of the field the energy spectrum of WS-states is discrete and, hence, they are localized states. We found the localization length of the honeycomb WS-states, which we define as the square root of the participation ratio (\ref{b6}), to grow in average as $1/F$, i.e., in the same way as for the square lattice. However, in a smaller scale the localization length shows large fluctuations that do not present in the case of square lattice. This observation motivates us to put forward a hypothesis about irregular (chaotic) nature of the honeycomb WS-states. This hypothesis is further supported by the irregular character of BO for irrational directions of the static field. 

The results of this work can be verified in laboratory experiments with cold atoms in honeycomb optical lattices \cite{Zhu07,Zhan12}, honeycomb photonic crystals \cite{Pele07,Szam11}, and microwave billiards with honeycomb array of scatterers \cite{Bitt10}. The common feature of these systems is that they offer direct visualization of the `wave function'. Another direction is ballistic conductivity of the graphene sheets. It is expected that the reported non-analytic angular dependence of WS-states could strongly affect the conductivity. We reserve the latter problem for future studies.

The author acknowledge support  of Russian Academy of Sciences through the SB RAS integration project  No.29 {\em Dynamics of atomic Bose-Einstein condensates in optical lattices} and the RFBR project No.12-02-00094 {\em Tunneling of the macroscopic quantum states}.

%%%%%%%%%%%%%%%%%%%%%%%%%%%%%%%%%%%%%

%%#############################################
%\begin{figure}
%\center
%%\includegraphics[height=9cm,clip]{fig2a.eps}
%\includegraphics[height=9cm,clip]{fig2c.eps}
%\caption{Wave function at $t=10T$ for $\theta=\pi-3$ and $F=1$. The figure shows population of the lattice sites by using specific encoding, where each site is substituted by the two-dimensional Gaussian multiplied by $|c_{l,m}|^2$.}
%\label{fig2}
%\end{figure}


\begin{thebibliography}{10}
\bibitem{remark1}
The triangular lattice can be mapped into a square lattice with additional diagonal hopping. Then, adopting the results of Ref.~\cite{Keck02}, one can prove that the energy bands a flat for all rational directions excluding $\theta=\pm\pi/6$ and $\theta=\pi/2$. %\cite{Bulg12}.
\bibitem{remark4}
In fact, these states are related to each other by symmetry transformation and, hence, have the same $P$.
\end{thebibliography}
\end{document}